\PassOptionsToPackage{table}{xcolor}
\documentclass[runningheads]{llncs}

\usepackage{amsmath,amssymb,amsfonts}
\usepackage{graphicx}
\usepackage{xcolor}
\usepackage{hyperref}
\hypersetup{
    colorlinks=true,
    linkcolor=blue,
    filecolor=magenta,      
    urlcolor=cyan,
}

\newcommand{\XHIDE}[1]{}

\title{IMPRESS: Improving Engagement in Software Engineering Courses through Gamification\thanks{
The IMPRESS project \url{https://impress-project.eu/} is funded by EU Erasmus{+} Programme, grant nr. 2017-1-NL01-KA203-035259. Duration: 2017-2020. Partners: 
  Open Univ. (NL), 
  Utrecht Univ. (NL), 
  Univ. Complutense Madrid (SP), 
  Univ. Passau (DE), 
  INESC-ID Lisbon (PT). 
The project is also partially funded by the 
  Fundac\~{a}o para a Ci\^{e}ncia e a Tecnologia (FCT) fund
  UID/CEC/50021/2019. 
}\thanks{This is a PREPRINT. The final version is published
in the proceedings of the 20th Int. Conf. on Product-Focused Software Process Improvement, LNCS 11915, 2019. This final version is available online at \url{https://doi.org/10.1007/978-3-030-35333-9_47}.
}
}

\author{Tanja E.J. Vos\inst{1}
\and I.S.W.B. Prasetya\inst{2}\orcidID{0000−0002−3421−4635}
\and Gordon Fraser\inst{3}
\and Ivan Martinez-Ortiz\inst{4}
\and Ivan Perez-Colado\inst{4}
\and Rui Prada\inst{5}
\and Jos\'e Rocha \inst{5}
\and Ant\'onio Rito Silva\inst{5}\orcidID{0000-0001-9840-457X}
}

\institute{Open Univeriteit Nederland
\and Utrecht University
\and Universit\"at Passau
\and Universidad Complutense de Madrid
\and INESC-ID and Instituto Superior T\'ecnico, Universidade de Lisboa
}

\begin{document}

\maketitle

\vspace{-5mm}
\begin{abstract}
Software Engineering courses play an important role for preparing students with 
the right knowledge and attitude for software development in practice. The implication is far reaching, as the quality of the software that we use ultimately depends on the quality of the people that make them. Educating Software Engineering, however, is quite challenging, as the subject is not considered as most exciting by students, while teachers often have to deal with exploding number of students. The EU project IMPRESS seeks to explore the use of gamification in educating software engineering at the university level to improve students' engagement and hence their appreciation for the taught subjects. This paper presents the project, its objectives, and its current progress.
\end{abstract}

\vspace{-8mm}

\section{Introduction}

\vspace{-2mm}
While our society increasingly depends on software for various aspects of civic, commercial and social life, software engineers struggle to ensure that software achieves the necessary high quality. The increasing complexity of modern software systems and the ever reducing time-to-marked further exacerbate the problem. Although the discipline of Software Engineering offers different techniques to ensure quality, programmers in practice are reluctant to engage with them, with detrimental effects on software quality.
The root of this situation lies in how software developers are educated. The focus tends to lie on the creative aspects of design and coding, whereas the more laborious and less entertaining necessities to assure the software's quality are neglected. This disengagement carries over to practice. This has to change: tomorrow software engineers need to be raised with appreciation of software quality, and quality assurance techniques need to become a natural aspect of software development, rather than a niche topic. Implementing the change, however, is not easy, as teachers have to motivate students through materials already branded as uninteresting. To help  teachers, the IMPRESS project seeks to explore the use of gamification, i.e., the application of game-design elements and game principles in non-gaming contexts, which has seen successful applications in other domains. This paper will present the project objective, the results so far, and a conclusion.

\vspace{-3mm}
\section{IMPRESS Objective}

\vspace{-3mm}
Although {\em gamificaton} is known to improve users' engagement and appreciation \cite{hamari2014does}, its application to Software Engineering is still limited. IMPRESS seeks to explore this  towards improving students' engagement and enthusiasm on topics traditionally considered as boring. The following focuses are chosen:

(1) Improving {\em in-class engagement} through gamified quizzes. Quizzes are an effective tool to set a course's pace. A cleverly setup quiz can trigger an engaging discussion, while gamification can stimulate wider engagement through competitive elements. A set of quizzes from selected topics will be developed within the project, along with tools to let others to develop more.
    
(2) Improving {\em out-class engagement} through educational games that can be played at home or in unguided lab sessions. We will focus on the subject of quality assurance ---a key subject, as pointed out earlier---, in particular in two key competences:  formalizing specifications and unit testing.
    
(3) Enhancing gamification with {\em story telling AI} for better emotional engagement and {\em advanded analytics} to provide insight on students' learning progress.

\vspace{-3mm}
\section{IMPRESS Innovations}

\vspace{-3mm}
This section presents the project progress so far.


\vspace{-4mm}
\subsubsection*{Keeping students on the move with quizzes.}


Quizzes have great potential as teaching tools. They can enrich the presentation of a course's content, and foster participation in the class subject. Tools like
\href{https://kahoot.com/}{Kahoot} prospered because of this. Quizzes can be used in a class to raise attention to particular issues, e.g. by showing to the students what they do not know, hence, supporting self-awareness of knowledge and make students more receptive to new information. Quizzes can also be used to support revision of knowledge, for example, as a summary in the end of the class, and to evaluate students.
\XHIDE{Teachers can use the results Quizzes to guide the discussion in class and to prepare materials for future classes. Theses results can give a good overview of the topics that students are struggling most.}
Outside the class, quizzes can be a good self assessing tool for students and enhancing their learning process by supporting self-regulation of learning and providing quick feedback about their current state of readiness on their subjects.

We have developed a web-based tool to reduce teachers' effort in preparing quizzes.
\XHIDE{, to allow them to more easily explore the potential of quizzes for teaching.} The tool,
available in a GitHub repository: \url{https://github.com/socialsoftware/as-tutor},
allows users to search through a repository of questions and quizzes, and create new quizzes by re-using and re-purposing the materials they find. 
The tool also supports automatic generation of quizzes on students' (or teachers') requests, e.g. classified according to a set of topics.
Produced quizzes can then be exported to gamified quiz tools, e.g. ARSnova, \url{https://arsnova.eu/}.
The repository currently contains over 600 questions and 80 quizzes, mostly on the subject of Software Architecture. A pilot in some of our courses is planned, after which the tool will be deployed open for the community.  We plan to extend the tool with automatic classification of questions (for more accurate automatic quizz generation) and generation of post-quizz feedback for both students and teachers on the students' learning progress. 

\XHIDE{
We will test the tool in a Software Architecture course at Instituto Superior T\'ecnico and open it to the community afterwards.
}

\XHIDE{
\begin{figure}
\begin{center}
\includegraphics[width=0.8\columnwidth]{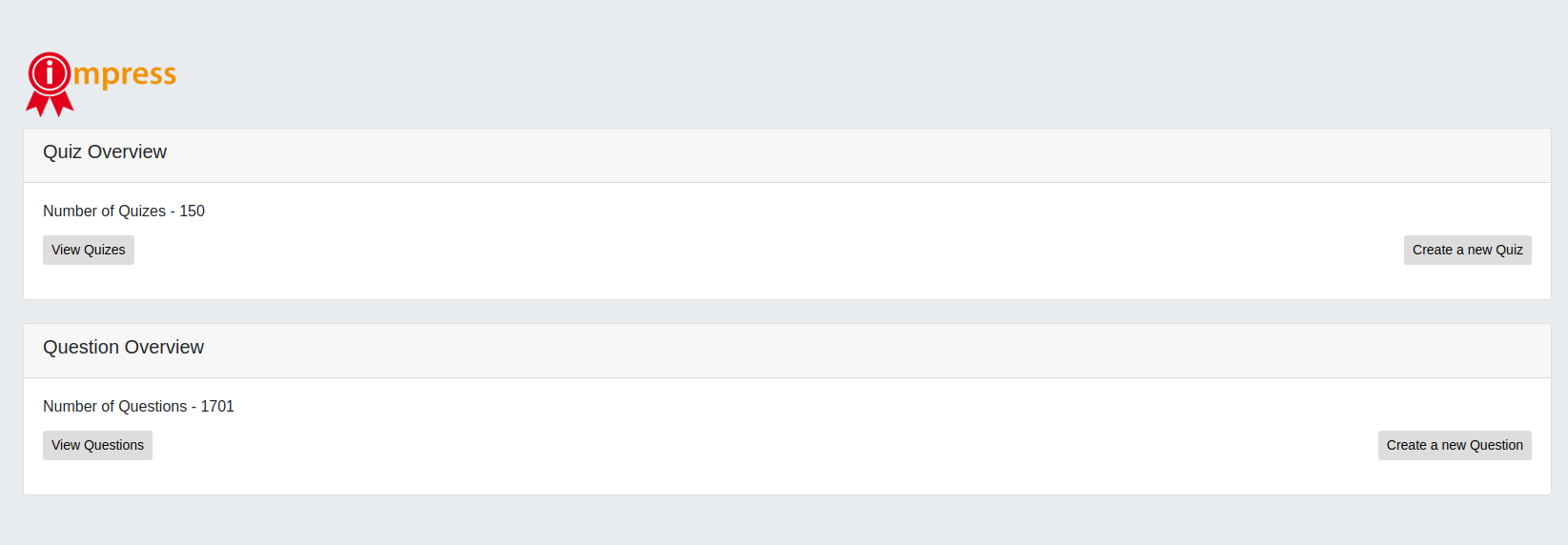} 
\end{center}
\vspace{-3mm}
\caption{\em IMPRESS Quiz and Question sharing and re-use site.}
\label{fig.quizsite}
\end{figure}
}

\XHIDE{
With this tool we want to make it possible to re-use World-wide existing set of questions for the Software Engineering related subjects. We will encourage software engineering teachers to share they experience and content to create a solid body of content to use, adapt and re-use. The repository has more than 1000 questions and 100 quizzes, at the moment, mostly from Software Architecture. We will test the tool in a Software Architecture course at Instituto Superior T\'ecnico and open it to the community afterwards.
}

\XHIDE{
Quizzes are a good base to support gamification. Replying to quizzes can be a task in a gamified system that grants points and progression to students/players. We are additionally working on our own quiz tool to run the quizzes and provide a personalized experience to the students by automatically suggesting questions related to the topics of studying with adequate difficult level.
}

\vspace{-5mm}
\subsubsection*{Training formalization skill with a game.}
\begin{figure}
\begin{center}
\includegraphics[scale=0.22]{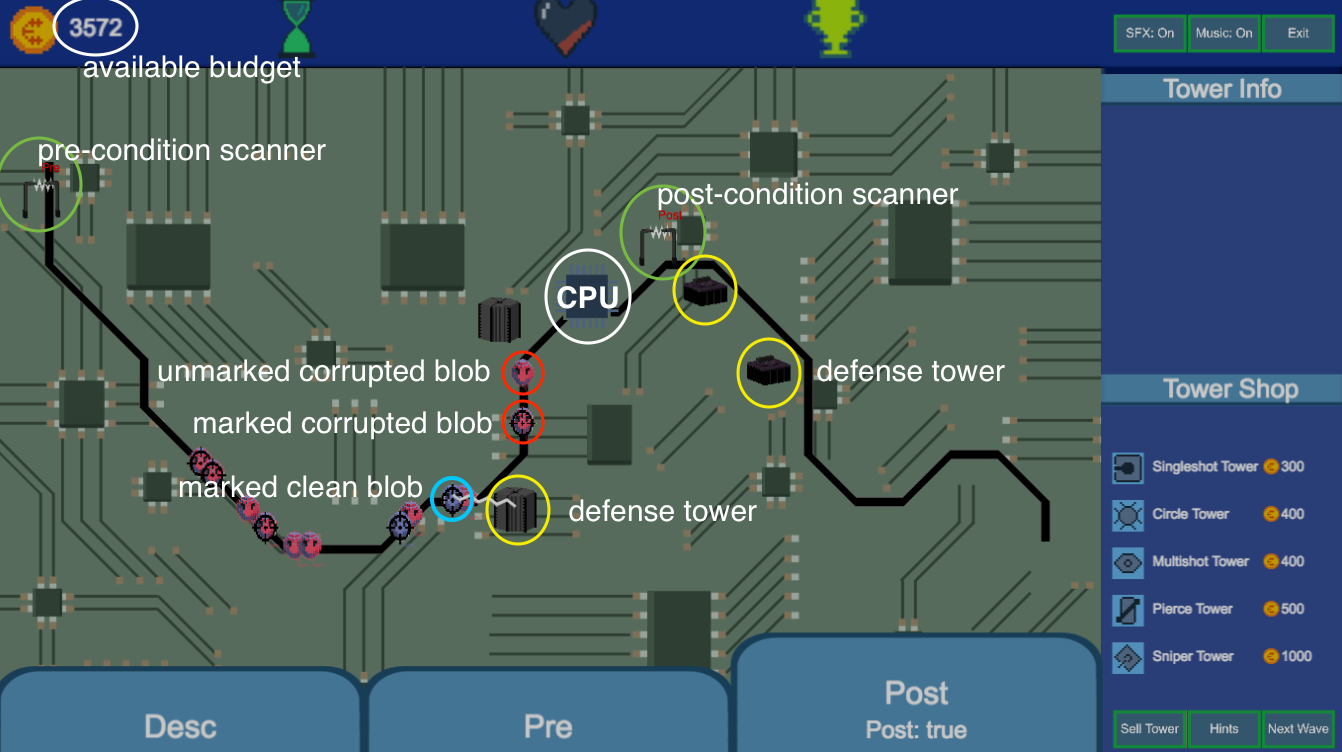} 
\end{center}
\vspace{-7mm}
\caption{\em A screenshot of FormalZ. The game is to defend the CPU in the middle of the circuit board.
The small red and blue blobs represent data coming to or leaving the CPU. Some of them might be corrupted. The user builds pre- and post conditions, and defense towers, trying to eliminate corrupted blobs. See also \cite{FormalZ}.}
\label{fig.gameshot1}
\end{figure}

Writing formal specifications is a skill that would greatly benefit students. Software with formal specifications can be verified, or at least tested, 
{\em automatically}, hence
greatly improving its correctness assurance. Unfortunately, this skill is often left underdeveloped. 
\XHIDE{
In the past, learning the skill is made difficult because students first 
need to learn sophisticated mathematical notations, e.g. Z \cite{xxx}. 
Fortunately, this is no longer the case  
since modern programming languages such as Java, C\#, or Haskell are often
expressive enough to allow programmers to write formal specifications natively 
and cleanly in these languages themselves. } 
The skill is not easy to master: 
it is easy to make mistakes, and training it 
can quickly become boring. 
In IMPRESS we experiment with a new game called FormalZ \cite{FormalZ} to train
the basic of writing formal specifications in the form of pre- and
post-conditions. Unlike existing Software Engineering
themed education games like Pex \cite{tillmann2011pex} and
Train-Director-B \cite{korevcko2015},
FormalZ takes a deeper gamification approach~\cite{boyce2014deep}, where 'playing' 
is given a more central role. After all, what makes games so engaging
is not merely the awarded scores and badges, but primarily the experience
of playing them. Fig. \ref{fig.gameshot1} shows a screenshot of FormalZ. 

\vspace{-5mm}
FormalZ also takes a {\em Constructionism} approach~\cite{papert1991constructionism}: just typing in formulas, which would be faster, is forbidden. Instead, the user  constructs formulas by dragging and connecting blocks of electronic hardware components. 
The Constructionism theory believes that humans learn by {\em constructing} knowledge, rather 
than by simply copying it from the teacher. 
Framing the knowledge in terms of
familiar physical objects, such as electronic components, plays a key role in this process,
because the learner already has knowledge on how they work~\cite{KafaiConstructionism05},
which the learner then uses 
to construct the new knowledge in his mind.
The theory was originally proposed by Papert and Harel~\cite{papert1991constructionism} and was e.g. used in the programming language LOGO for teaching programming to children.

The initial reaction from our students have been encouraging \cite{FormalZ}, but more studies are needed to investigate the actual impact
on the game's learning goal. 

\vspace{-3mm}
\subsubsection*{Teaching software testing through a competitive game.}
A further challenging activity in software engineering practice as well as education is testing a program for errors. In IMPRESS we explore improving the education of testing using Code Defenders, a game intended to engage students in the context of a Java object-oriented class under test and its test suite. In the game, \emph{attackers} aim to introduce artificial bugs (``mutants'') into the class under test that reveal weaknesses in the test suite, while \emph{defenders} aim to improve the test suite by adding new tests. If a mutant program produces a different output for a test than the original program, then that mutant is detected by the test, and the defender who wrote the test scores points. If a mutant is not detected by any tests, then the attacker scores points. The number of points a mutant is worth depends on the number of tests it ``survives'', which further encourages players to create as subtle as possible mutants, and as strong as possible tests.

\begin{figure}
\centering
\includegraphics[width=0.7\columnwidth]{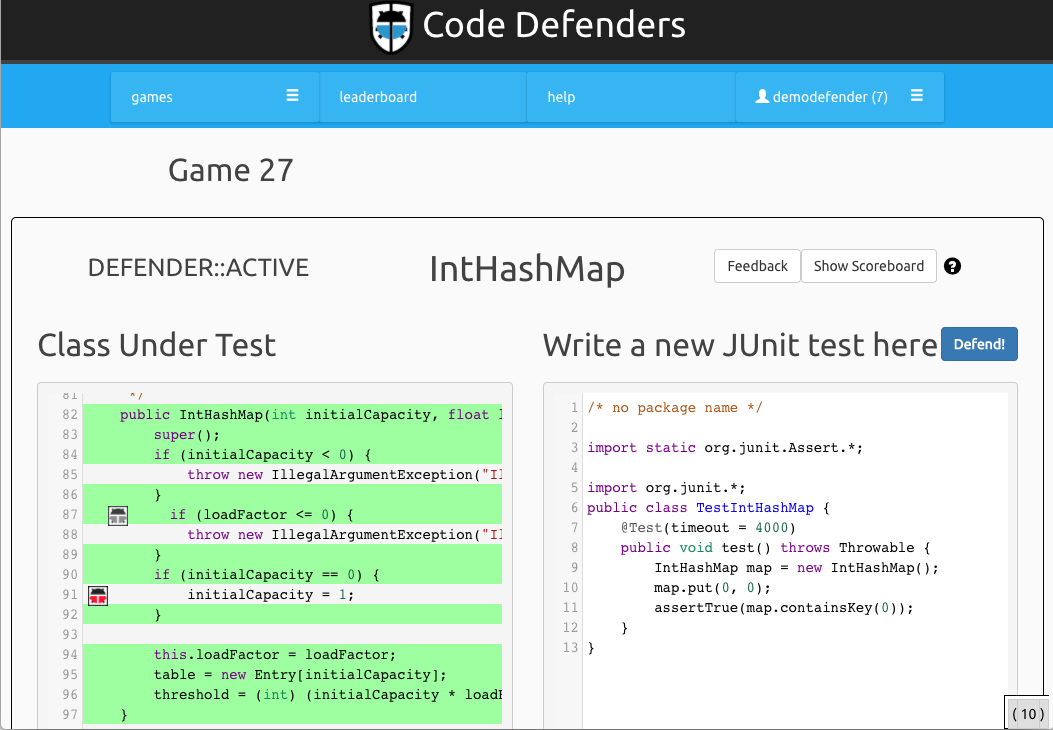}
\vspace{-3mm}
\caption{\label{fig:codedefenders}\em The Code Defenders game in action: the source code under test is shown together with information about the coverage of the existing tests and bugs location. Defenders add new tests, while attackers create ``mutants'' by editing the source code directly.}
\end{figure}

Code Defenders is implemented as a web-based game
and is played by teams of students. The players are shown the source code of the Java class under test, with color highlighting to indicate the coverage of the defenders' test suite, and with bug-icons labelling the locations and status of the attackers' mutants. Attackers create mutants by editing the source code of the Java class, and defenders write JUnit tests using a code editor. A scoreboard breaks down the game's current score for each team and player.

We have studied player behavior in detail~\cite{rojas2017code} and shown that players enjoy writing tests in the game more than as a regular developer activity. We have also applied Code Defenders in class and designed a software testing undergrad course around it~\cite{fraser2019gamifying}. Initial evaluation results suggest that Code Defenders supports students in achieving their learning objectives.

\vspace{-3mm}
\subsection{Advanced analytics}

\vspace{-2mm}
We have extended the analytics platform from the H2020 RAGE project\footnote{GitHub repository: https://github.com/e-ucm/rage-analytics} to adequate its functionalities to IMPRESS' needs, in particular to support different types of analytics generating educational activities \cite{ivan2019}. These new developments allowed two approaches for analytics integration: {\em light} and {\em deep} integration.

Often, educational tools (like Kahoot!) provide a report that summarizes students interaction to some extent. 
In light integration the underlying educational tool it is not modified at all (e.g. because  modification is not possible). RAGE Analytics is simply used on  available analytics provided by the educational tool, e.g. to provide better or uniform visualisation across multiple tools. 

\begin{figure}
\begin{center}
\includegraphics[scale=0.35]{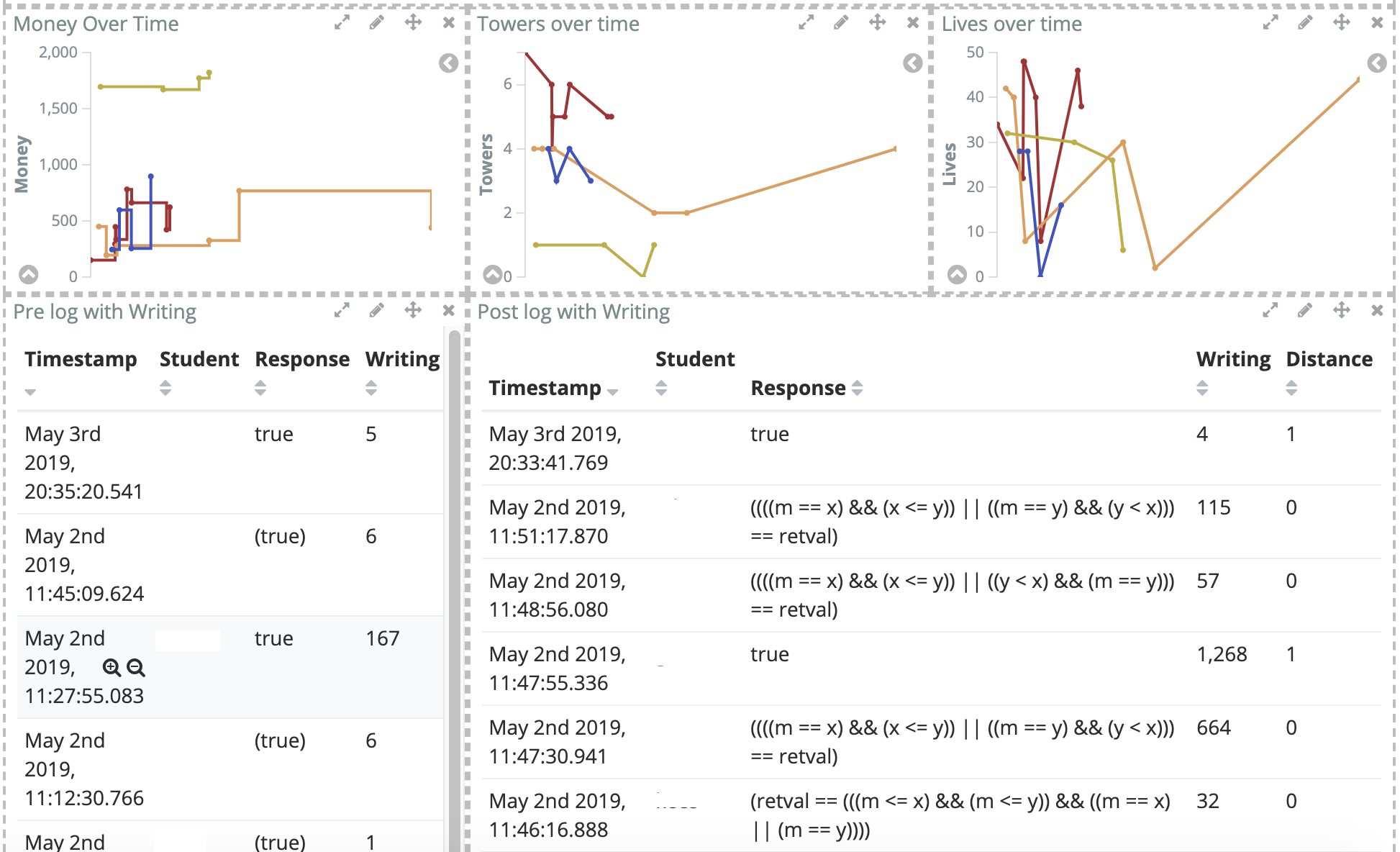} 
\end{center}
\vspace{-6mm}
\caption{\em FormalZ analytics main dashboard.}
\label{fig.formalZdashboard}
\end{figure}

In deep integration, the developers of the education tool need to integrate a “tracker” \cite{ivanPC2018-02} into the tool, used to send out the user interaction information. As such, this approach can provide more fine grained analytics and to provide it live and is therefore the recommended integration approach.
This was the approach selected for integration of the FormalZ game with RAGE Analytics, allowing us to collect all students interactions and to show them graphically to teachers, near real-time, in a single dashboard (Fig.~\ref{fig.formalZdashboard}). The analytics can also show how the students evolve their solutions, to give insight on their mental process in constructing the solutions.

Having all analytics in one place allowed us to provide an additional capability for teachers that want to have analytics of multiple heterogeneous activity (e.g. to track student progress during a longer period). This is facilitated through configurator to perform simple operations and weight of activities, so they can build new variables that can be included in class level dashboards \cite{ivanPC2018}.

\subsection{AI in IMPRESS}

One of the use of AI for teaching is the generation and adaptation of learning content\cite{brisson2017AIpersonalization}. We are currently working on an AI module to create personalization features of the previously mentioned quiz tool we developed. It will work with the data that will be stored by the students performance on the quizzes to define student profiles and choose the best quizzes to enrich their learning experience.

AI can also improve the learning experience by adding a storytelling layer to the content.
Stories are common in games and support meaning making and emotional engagement that foster learners motivation and learning\cite{ohler2013storytelling}. We are developing storytelling components for the Code Defenders and FormalZ games by using the FAtiMA toolkit\footnote{https://fatima-toolkit.eu/}\cite{mascarenhas2018fatimatoolkit}. Our approach is to put the challenges presented by the games into a narrative, by including a character in the game that will talk to the players contextualizing the challenge that is given to the player(s) and presenting feedback on the performance. The toolkit facilitates the creation of such characters including mechanisms for the generation of personality and emotional responses, an authoring tool for character's behaviour, and integration through a REST API.

\vspace{-3mm}
\section{Conclusion}

\vspace{-3mm}
While the importance of Software Engineering courses is  well acknowledged, creating engaging Software Engineering courses is very challenging. Much can be improved through innovative use of modern technology. 
Along this line, IMPRESS has contributed innovations in gamification, and more can be expected before the project ends in 2020. Ultimately though, energizing Software Engineering education is not a challenge that a single project like IMPRESS can solve on its own. Community, and Industry, should also own the problem and commit to solving it.

\bibliographystyle{splncs04}
\bibliography{references.bib}

\begin{thebibliography}{10}
\providecommand{\url}[1]{\texttt{#1}}
\providecommand{\urlprefix}{URL }
\providecommand{\doi}[1]{https://doi.org/#1}

\bibitem{boyce2014deep}
Boyce, A.K.: Deep Gamification: Combining Game-based and Play-based Methods.
  Ph.D. thesis, North Carolina State Univ. (2014)

\bibitem{brisson2017AIpersonalization}
Brisson, A., Pereira, G., Prada, R., Paiva, A., Louchart, S., Suttie, N., Lim,
  T., Lopes, R.A., Bidarra, R., Bellotti, F., Kravcik, M., Oliveira, M.F.:
  Artificial intelligence and personalization opportunities for serious games.
  In: Proc. of the 8th Artificial Intelligence and Interactive Digital
  Entertainment Conf. (2012)

\bibitem{fraser2019gamifying}
Fraser, G., Gambi, A., Kreis, M., Rojas, J.M.: Gamifying a software testing
  course with code defenders. In: Proceedings of the 50th ACM Technical
  Symposium on Computer Science Education. pp. 571--577. ACM (2019)

\bibitem{hamari2014does}
Hamari, J., Koivisto, J., Sarsa, H., et~al.: Does gamification work? --a
  literature review of empirical studies on gamification. In: 47th Hawaii
  International Conference on System Sciences (2014)

\bibitem{KafaiConstructionism05}
Kafai, Y.B.: The Cambridge Handbook of the Learning Sciences, chap.
  Constructionism. Cambridge University Press (2005)

\bibitem{korevcko2015}
Kore{\v{c}}ko, {\v{S}}., Sor{\'a}d, J.: Using simulation games in teaching
  formal methods for software development. In: Innovative Teaching Strategies
  and New Learning Paradigms in Computer Programming, pp. 106--130. {IGI}
  Global (2015)

\bibitem{ivan2019}
{Martínez-Ortiz}, I., {Pérez-Colado}, I., {Rotaru}, D.C., {Freire}, M.,
  {Fernández-Manjón}, B.: From heterogeneous activities to unified analytics
  dashboards. In: IEEE Global Engineering Education Conference (EDUCON) (2019)

\bibitem{mascarenhas2018fatimatoolkit}
Mascarenhas, S., Guimar{\~a}es, M., Prada, R., Dias, J., Santos, P.A., Star,
  K., Hirsh, B., Spice, E., Kommeren, R.: A virtual agent toolkit for serious
  games developers. In: Proc. Conf. on Computational Intelligence and Games
  (CIG). IEEE (2018)

\bibitem{ohler2013storytelling}
Ohler, J.B.: Digital Storytelling in the Classroom: New media pathways to
  literacy, learning, and creativity. Corwin Press (2013)

\bibitem{papert1991constructionism}
Papert, S., Harel, I.: Constructionism. Ablex Publishing (1991)

\bibitem{ivanPC2018-02}
{Perez-Colado}, I., {Alonso-Fernandez}, C., {Freire}, M., {Martinez-Ortiz}, I.,
  {Fernandez-Manjon}, B.: Game learning analytics is not informagic! In: 2018
  IEEE Global Engineering Education Conference (EDUCON) (2018)

\bibitem{ivanPC2018}
{Perez-Colado}, I.J., {Rotaru}, D.C., {Freire-Moran}, M., {Martinez-Ortiz}, I.,
  {Fernandez-Manjon}, B.: Multi-level game learning analytics for serious
  games. In: 10th Int. Conf. on Virtual Worlds and Games for Serious
  Applications (VS-Games) (2018)

\bibitem{FormalZ}
Prasetya, I.S.W.B., Leek, C.Q.H.D., Melkonian, O., Tusscher, J.t., van Bergen,
  J., Everink, J.M., van~der Klis, T., Meijerink, R., Oosenbrug, R., Oostveen,
  J.J., van~den Pol, T., van Zon, W.M.: Having fun in learning formal
  specifications. In: Proc. 41st Int. Conf. on Software Engineering ({ICSE}).
  IEEE (2019)

\bibitem{rojas2017code}
Rojas, J.M., White, T.D., Clegg, B.S., Fraser, G.: Code defenders:
  crowdsourcing effective tests and subtle mutants with a mutation testing
  game. In: Proc. 39th Int. Conf. on Software Engineering. IEEE Press (2017)

\bibitem{tillmann2011pex}
Tillmann, N., de~Halleux, J., Xie, T.: Pex for fun: Engineering an automated
  testing tool for serious games in computer science. Tech. rep.,
  MSR-TR-2011-41 (2011)

\end{thebibliography}

\end{document}